\documentclass[a4paper,11pt]{article}
\usepackage[dvips]{graphicx}
\usepackage{amssymb}
\usepackage{amsmath}

\usepackage{cite}

\begin{document}

\title{Testing extra dimensions with boundaries using Newton's law modifications}
\author{
V.K. Oikonomou $^{1}$\,\thanks{voiko@physics.auth.gr}\,\,
and K. Kleidis $^2$\,\thanks{kleidis@teiser.gr}\\
\emph{$^1$Department of Informatics and Communications} \\
\emph{Technological Education Institute of Serres}\\
\emph{62124 Serres, Greece}\\
\emph{$^2$Department of Mechanical Engineering, Technological}\\
\emph{ Education Institute of Serres}\\
\emph{62124 Serres, Greece} \\
} \maketitle

\begin{abstract}
Extra dimensions with boundaries are often used in the literature, to provide phenomenological
models that mimic the standard model. In this context, we explore possible modifications to Newton's law due to the
existence of an extra-dimensional space, at the boundary of which
the gravitational field obeys Dirichlet, Neumann or mixed boundary conditions. We focus on two types of
extra space, namely, the disk and the interval. As we prove, in order to
have a consistent Newton's law modification (i.e., of the Yukawa-type), some of the extra-dimensional spaces that have been used in the literature, must be ruled out.
\end{abstract}

\maketitle

\section{Introduction}

One of the most challenging perspectives in contemporary physics
is the unified description of all the fundamental forces in nature
within a self-consistent theoretical framework. In this context, a
vast amount of theoretical work has been carried out during the
last sixty years.

A successful effort to incorporate all interactions into a unified
scheme is  M-theory, together with its low energy limits, the Type
IIA and Type IIB string theories, the $E_8\times E_8$ heterotic
string theories
\cite{Polchinski1998,Leontaris2008,Leontaris2009,Johnson2003}, and
so forth. These scenarios suggest that spacetime possesses
additional compact spatial dimensions. Unfortunately, the
corresponding compactification scale is much lower than the scales
that can be experimentally examined. There are also
string-inspired extensions of the four-dimensional quantum field
theory, including large extra dimensions \cite{Antoniadis1998}, in
which the compactification scale is of the order of T$eV$. Such
models appear to incorporate consistently the standard-model
phenomenology \cite{Kribs2006} and, therefore, they could set a
path towards higher-dimensional extensions of other
quantum-field-theory areas (see
\cite{Buchbinder1988,Odintsov1988,Lavrov1988,Buchbinder1989,Kirsten1991,Kirsten1992,Kirsten1993,Elizalde2003,Elizalde2004,Elizalde2008,Oikonomou2009a,Oikonomou2009b,Oikonomou2010c,Oikonomou2010a},
for an extensive though incomplete list). Some of these extensions
could be verified, even experimentally \cite{Antoniadis1998}. In
this context, it is expected that the existence of extra
dimensions can be further scrutinized in the Large Hadron
Collider, now operating at CERN. A potential experimental
verification of the extra dimensions existence is quite
challenging and, perhaps, the already developed experiments could
be enriched also with others, corresponding to alternative types
of approach \cite{Kazanas2001,Domokos2003,Oikonomou2007}.

A theoretical concept that could be verified experimentally in the
near future, is the potential modification of gravity in three
dimensions, due to the existence of extra dimensions. In fact,
many theoretical studies have already been accounted for such a
perspective \cite{Floratos1999, Kehagias2000,Kehagias2001,
Oikonomou2008,Floratos2010,Oikonomou2010b}. According to these
models, the gravitational potential in three dimensions (Newton's
law) can be modified, due to a compact $n$-dimensional extra
space, thus resulting in Yukawa type corrections, which are of the
form:
\begin{equation} \label{1}
V_3(r) = - G_N \frac{M M'}{r} \left ( 1 + a e^{-
\frac{r}{m_1}}\right )
\end{equation}
where $r$ is the radial coordinate in three dimensions, $G_N$ is
Newton's universal constant of gravitation in four spacetime
dimensions, $a$ is the (dimensionless) strength of the gravitational field and
$m_1$ is the length scale involved. According to
Eq. (\ref{1}), the correction to the Newtonian potential,
$\frac{1}{r}$, is of the Yukawa type, $a e^{- \frac{r}{m_1}}/r$.
In this context, $m_1 = m^{-1}$, where $m$ is the mass of the
lightest Kaluza-Klein state (which is the leading-order mode). At
distances relatively-close to the length-scale of the extra
dimensions, $R$, macroscopic modifications are expected, both to
the strength and to the range of the gravitational force and the
three-dimensional gravitational potential changes
\cite{Floratos1999,Kehagias2000,Oikonomou2008,Floratos2010,
Oikonomou2010b}. A possible modification of Newton's law, due to
the existence of extra dimensions, could be revealed by several
types of experiments \cite{Kribs2006}. As a consequence, there are
also several bounds constraining the size and number of the extra
dimensions \cite{Kazanas2001,Olive2003,Vergados2006,Oikonomou2007,
Hooper2009,Floratos2010}. Nevertheless, as far as the modification
of Newton's law is concerned, the strongest constraint comes from
the E\"{o}tvos and Cavendish-type of experiments, which strongly
restrict the parameters $a$ and $m_1$ \cite{Kribs2006}. Their
results are consistent with the theoretical predictions of
Newtonian gravity down to 200 microns \cite{Hoyle2004} (2004
result), while more recent results suggest that Newtonian gravity is valid
down to $56$ microns \cite{Hoyle2006}.

In this letter, we explore possible Yukawa-type modifications to
the gravitational potential in three dimensions, due to the
existence of extra-dimensional spaces, at the boundary of which
the extra dimensional component of the gravitational field obeys
Dirichlet, Neumann or mixed boundary conditions. We consider two
types of extra space, a disk with compact radius $R$ and the interval with length $ 2 \pi R$. The aforementioned spaces have
been frequently used in the extra-dimensional phenomenology
literature. The interval, in particular, has been used in
\cite{csaba} in order to break gauge symmetries. Various gauge-symmetry-breaking patterns arise using combinations
of Dirichlet and Neumann boundary conditions (these boundary
conditions are quite similar to orbifold boundary conditions).
These theories can accommodate the standard model without
introducing the Higgs field. Another perspective of the
interval is studied in \cite{Diego2008}. On the same footing, the
Dirichlet disk and the Neumann disk have already been exploited
\cite{Dudas2006,DudasRubakov2006}, offering interesting
phenomenology to the extra-dimensional extensions of the standard
model. These models describe a scalar theory built on a
six-dimensional spacetime with a Dirichlet disk (of radius $R$) as
the extra dimensional space. The radius $R$ serves as the
intermediate scale. The Dirichlet disk is used in order that, the
singularity involved, to be resolved. Such models also appear in
string theory \cite{Dudas2006}, where massless or very light
degrees of freedom emerge upon turning on small couplings, in a
framework which possesses only heavy degrees of freedom. Our
findings constrain the set of the allowed boundary conditions that
the above models use. Moreover we exclude some spaces because
these provide unphysical Newton's law corrections.
\footnote{Assuming that the extra dimensional component of the
gravitational field, obeys the same boundary conditions as the
rest of the fields in the theory.}

This paper is organized as follows: In Section 2, we present the
basic formulas that yield Yukawa-type corrections to Newton's
gravitational potential in three dimensions, due to the existence
of extra spatial dimensions. In this context, we explore possible
modifications to Newton's law arising from a higher-dimensional spacetime with a disk as the extra space, at
the boundary of which the gravitational field obeys either Dirichlet or Neumann boundary conditions. In Section 3, we analyze the interval involving all the potential combinations of boundary conditions. We generalize to the
cases of two- and three-dimensional intervals and we
compare our results with those corresponding to other well-studied
extra-dimensional spaces (such as the 2-sphere, the 3-torus etc.).
In addition, we perform a brief study of the mixed (Robin) boundary conditions.
In Section 4, we check the consistency of our results with those of the experimental methods used to test the validity of Newton's law. Finally, we present our conclusions in
Section 5.

\section{Gravitational potential and extra dimensions-the case of a disk}

Here, we recall the most important formulas related to the Yukawa-type
corrections of the gravitational potential in three dimensions,
due to the existence of extra spatial dimensions (for a detailed analysis see, e.g., \cite{Kehagias2001}).

We consider a background of the form $M^4 \times M^n$, where $M^n$
stands for a $n$-dimensional compact manifold (no topology is
specified) and $M^4$ is the four-dimensional Minkowski spacetime
used in contemporary particle physics. We assume that there exists
a complete set of orthonormal harmonic functions, $\Psi_m (x)$,
defined on $M^n$. These harmonic functions are eigenfunctions of
the $n$-dimensional Laplace-Beltrami operator $\Delta_n$ defined
on $M^n$, with eigenvalues $\mu_m^2$
\begin{equation}\label{6} - \Delta_n \Psi_m = \mu_m^2 \Psi_m \: . \end{equation}

Provided that the length-scale of the compact dimensions, as
compared to the radial distance (from the source) of the
three-dimensional space, is small, the Newton's law in three
dimensions results in (for a detailed analysis consult
\cite{Kehagias2001}),
\begin{equation}\label{13} V_3 (r) = - \frac{G_N M}{r} \sum_m \Psi_m^* (0) \Psi_m
(0) e^{- \vert \mu_m \vert \: r} . \end{equation} As we have
already mentioned, Eq. (\ref{13}) is valid for large values of
$r$, as compared to the internal dimensions radii, and, in fact,
it can be expressed in the form \begin{equation}\label{14} V_3 (r)
= - \frac{G_N M}{r} \sum_m d_{m} e^{- \vert \mu_m \vert \: r} ,
\end{equation} where $d_m$ denotes the degeneracy of the eigenvalue $\mu_m$.
Eq. (\ref{14}) is particularly useful to those cases where the
representative eigenfunction is too difficult to be found. This is
due to the fact that the eigenvalues $\mu_m$ depend on the
irreducible representation of the gravitational field and not only
on the particular representative $\Psi$ \cite{Kehagias2001}.

\subsection{The Dirichlet disk as an extra-dimensional space}

Consider that the extra-dimensional space is a Dirichlet disk of
radius $R$. Upon consideration of the aforementioned analysis, the
overall gravitational potential is written in the form
\begin{equation}\label{16} V_{3+2} = \sum_m \Phi_m (r) \Psi_m (x) \: , \end{equation} where
$\Psi_m$ represents the extra-dimensional counterpart of the
gravitational potential, corresponding to harmonic functions on a
two-dimensional disk, at the boundary of which they obey Dirichlet
boundary conditions, i.e., $\Psi_m (R,\phi) = 0$. The disk is
parameterized by the radius $r$, with $0\leq r \leq R$ and by the
angle $\phi$, with $0\leq \phi \leq 2\pi$. The solutions to the
Laplace equation (\ref{6}) reads,
\begin{equation}\label{radial}
\Psi_m(r,\phi)=J_m(\frac{x_{mn}}{R}r)e^{im\phi}
\end{equation}
In this case, according to the solution of the eigenvalue equation
(\ref{6}), the gravitational potential in three dimensions reads
\begin{equation}\label{17} V_3 (r) = - \frac{G_N M}{r} \sum_{m,n} J_m^* (0) J_m
(0) e^{- \vert x_{mn} \vert \: \frac{r}{R}} \: , \end{equation}
where $J_m$ are the Bessel functions of the first kind and
$x_{mn}$ are their roots.\footnote{Note that, in Eq. (3), the value $x=0$ corresponds to $r=0$ and $\phi=0$, using the
disk parametrization.} In view of the particular special functions
properties, the only non-zero term of Eq. (\ref{17}) is the one
containing $J_0 (0) = 1$, since, for $n \neq 0$, $J_n (0) = 0$.
Therefore, Eq. (\ref{17}) results in
\begin{equation}\label{18} V_3 (r) \simeq -\frac{G_N M}{r} \sum_n
e^{- \vert x_{0n} \vert \: \frac{r}{R}} \: . \end{equation} The
first root of $J_0$ is $x_{01} = 2.40483$ \cite{Abramowitz1970},
hence, in this case, the gravitational potential reads
\begin{equation}\label{19} V_3 (r) \simeq - \frac{G_N M}{r} e^{-
2.40483 \: \frac{r}{R}} \: . \end{equation} We kept only the
leading order corrections to Newton's law (the next to leading
order terms are exponentially suppressed, and therefore are
subdominant terms). Eq. (\ref{19}) implies a rather unexpected
result, in the sense that a compact space with radius in the
sub-millimeter scale should yield the four-dimensional Newton's
law plus a small correction of the Yukawa-type form. Hence, although it is interesting, it appears that the
Dirichlet disk cannot serve as an extra dimensional space, since
it results in an unexpected form of Newton's law, namely that of
the Yukawa type. Consequently, the extra-dimensional theories with the
Dirichlet disk as extra space must be ruled out.

\subsection{An internal disk with Neumann boundary}

Now, we explore the modifications to Newton's gravitational
potential in three dimensions in the case where the extra
two-dimensional disk obeys Neumann boundary conditions. The Neumann
disk was used in \cite{Dudas2006}, along the same lines as
those in the Dirichlet case described in the previous section. In
the Neumann case, the harmonic function $\Psi (x)$ obeys
$\frac{\mathrm{d} \Psi}{\mathrm{d} x} \vert_{x = R} = 0$ and the
gravitational potential in three dimensions is written in the form
\begin{equation}\label{23} V_3 (r) = -\frac{G_N M}{r} \sum_{m, n}
J_m^* (0) J_m (0) e^{- \vert x_{mn}^{\prime} \vert \: \frac{r}{R}}
\: , \end{equation} where $x_{mn}^{\prime}$ are the roots of the
Bessel function's derivative $J_m^{\prime} (x)$. The only non-zero
term involved in the sum on the rhs of Eq. (\ref{23}) is the one containing
$J_0 (0) = 1$ and, therefore, $V_3 (r)$ reads
\begin{equation}\label{24} V_3 (r) \simeq - \frac{G_N M}{r} \sum_n
e^{- \vert x_{0n}^{\prime} \vert \: \frac{r}{R}} \: .
\end{equation} The first root of the derivative of the Bessel
function involved is $x_{01}^{\prime} = 0$, while, the
corresponding second root (which is the lowest Kaluza-Klein mode)
is $x_{02}^{\prime} = 3.8317$. Subsequently, keeping the lowest
Laplacian eigenvalues $x_{01}$ and $x_{02}$, Eq. (\ref{24})
results in
\begin{equation}\label{25} V_3 (r) \simeq - \frac{G_N M}{r} \left ( 1 + e^{-
3.8317 \frac{r}{R}} \right ) \: . \end{equation} In the above
equation we kept only the leading order corrections to Newton's
law (which don't include subdominant terms). We observe that, in
this case, due to the properties of the harmonic functions
attributed to the internal disk (with Neumann boundary), the
three-dimensional gravitational potential acquires the usual form
of Newton's law, $\frac{1}{r}$, plus a Yukawa-type correction,
reminiscent to compact extra dimensions. Thereupon, a
two-dimensional internal space compactified on a disk with Neumann
boundary, could result in the Yukawa-type modification of Newton's
gravitational potential, in contrast to the corresponding
Dirichlet case.

\section{The Interval}

In this section we study another extra-dimensional space, that has
been used in several multi-dimensional extensions of the standard
model \cite{csaba,Diego2008}. It is conceptually very simple,
corresponding to an one-dimensional interval of length $2 \pi R$,
denoted as $I = [0, 2 \pi R]$. The spacetime structure is of the
form $M^4 \times I$, with $M^4$ being the four-dimensional
Minkowski spacetime. In this article, we are not interested in the
wealth of phenomenological features that this specific model may
offer \cite{csaba,Diego2008}. Instead, we explore the way that the
three-dimensional gravity is modified. The literature
distinguishes four cases with respect to the boundary conditions
admitted at $x = 0$ and at $x = 2 \pi R$, namely,
Dirichlet-Dirichlet (D-D), Dirichlet-Neumann (D-N),
Neumann-Dirichlet (N-D), and Neumann-Neumann (N-N) \cite{csaba}.
It is easy to see that the gravitational corrections of the D-N
and N-D extra spaces are unphysical, since there is no zero mode
corresponding to these boundary conditions. Indeed the eigenvalues
of the Laplace-Beltrami operator are $n+1/2$, with
$n=0,1,2,3,...$. Hence, we have a similar situation to the
Dirichlet disk we studied earlier. On the other hand, the D-D case
also yields unphysical results, since
\begin{equation}\label{eq}
V_3(r)\sim\sum_n \sin(0)^2e^{-\lambda_nr/R}
\end{equation}
which is zero (with $\lambda_n$ the D-D eigenvalues). The only
case that predicts Yukawa-type corrections is the N-N case.
Accordingly, the harmonic function, $\Psi (x)$, of the
gravitational potential along the extra spatial dimension $(x)$
obeys \begin{equation}\label{27} \frac{\mathrm{d} \Psi}{\mathrm{d}
x} \vert_{x = 2\pi R} = 0 \: , \, \, \, \frac{\mathrm{d}
\Psi}{\mathrm{d} x} \vert_{x = 0} = 0 \end{equation} and the
corresponding eigenfunctions are of the form
\begin{equation}\label{28} \Psi (x) \simeq \cos \left (
\frac{n{\,}x}{2R} \right ) \: , \, \, \, \, \, n = 0, \: 1, \: 2,
\ldots \end{equation} The correction to the three-dimensional
gravitational potential is given by,
\begin{equation}\label{29} V_3 (r) = - \frac{G_N M}{r} \sum_{n} d_n e^{- n \:
\frac{r}{2R}} \: , \end{equation} with $d_n$ being the degeneracy
of the eigenvalues of the Laplace-Beltrami operator along the
extra dimension. Once again, keeping the leading-order corrections
to Newton's law (omitting subdominant terms), we obtain
\begin{equation}\label{30} V_3 (r) = - \frac{G_N M}{r} \left ( 1 +
e^{- \frac{r}{2R}} \right ) \: . \end{equation} The generalization
of the interval to two and three dimensions\footnote{The two
dimensional space is of the form $M^4 \times I\times I$, while the
three dimensional is $M^4 \times I\times I\times I$.} (the lengths
of the intervals are considered equal) is straightforward and the
potential equals to:
\begin{align}\label{equations}
& V_3 (r) = - \frac{G_N M}{r} \left ( 1 + 2e^{-
\frac{r}{2R}}\right )(\mathrm{Two{\,}Dimensional{\,} Interval})
\\ \notag &
V_3 (r) = - \frac{G_N M}{r} \left ( 1 + 3e^{- \frac{r}{2R}}\right
) (\mathrm{Three{\,}Dimensional{\,} Interval})
\end{align}
It is assumed that the extra dimensional component of the field, obeys Neumann
boundary conditions at all boundaries.

Summarizing, an extra space with Neumann
boundaries could yield Yukawa-type corrections to Newton's law, in
contrast to the extra spaces with Dirichlet, or combinations of
Dirichlet and Neumann boundary conditions. This result is quite important, since it
rules out a large number of extra-dimensional spaces.

\subsection{Interval with Robin Boundaries}

Now, we consider that, at the first boundary of the interval, $x=0$, the extra-dimensional component of the gravitational
field, $\Psi$, obeys Robin boundary conditions \cite{saharian1}
of the form,
\begin{equation}\label{28vb}
(1+\beta_1\partial_x)\Psi (x)=0
\end{equation}
and, similarly, on the second boundary, $x = 2 \pi R$,
\begin{equation}\label{29dfgtg}
(1+\beta_2\partial_x)\phi (x)=0 \: ,
\end{equation}
where $\beta_{1,2}$ are arbitrary constants. Robin boundary conditions
are known as providing conformal invariance for field theories
\cite{saharian1}, and have been used in phenomenological models
such as those of \cite{csaba}.

We shall give a general rule that must be obeyed in
order to obtain the expected Yukawa type behavior of the
gravitational corrections. Therefore we must address the spectral
problem, subject to the Robin boundaries. The roots (which we
denote $y_n$) of the equation,
\begin{equation}\label{30df}
(1-b_1b_2y_n^2)\sin y_n-(b_1+b_2)y_n\cos y_n=0
\end{equation}
where $b_i=\beta_i /2\pi R $, are the eigenvalues of the extra
space's Laplace-Beltrami operator, with the gravitational field
obeying Robin boundary conditions \cite{saharian1}. If we denote
the corresponding eigenfunctions $\varphi_n(x)$, then in order to
have the Yukawa type corrections (and to avoid unphysical results)
the eigenvalues must have a zero mode and the eigenfunctions must
be non zero at $x=0$, namely:
\begin{equation}\label{eigenval}
\varphi_n(0)\neq 0.
\end{equation}
Then, the corrections of the gravitational potential take the
form,
\begin{equation}\label{robin}
V(r) = - \frac{G_N M}{r} \left ( 1 + \sum_{n\neq
0}\varphi(0)^*\varphi(0)e^{-y_nr}\right )
\end{equation}
which is the physically accepted (and formally expected) behavior
stemming from a compact extra space. It is obvious that the cases
D-D, N-D, D-N, and N-N are subcases of the Robin boundary
conditions.

\section{Analysis and Comparison of the Results}

In this section we check upon the consistency of our results
with the experimental ones on the validity of Newton's law.
\begin{figure}[ht!]
\begin{minipage}{18pc}
\includegraphics[width=20pc]{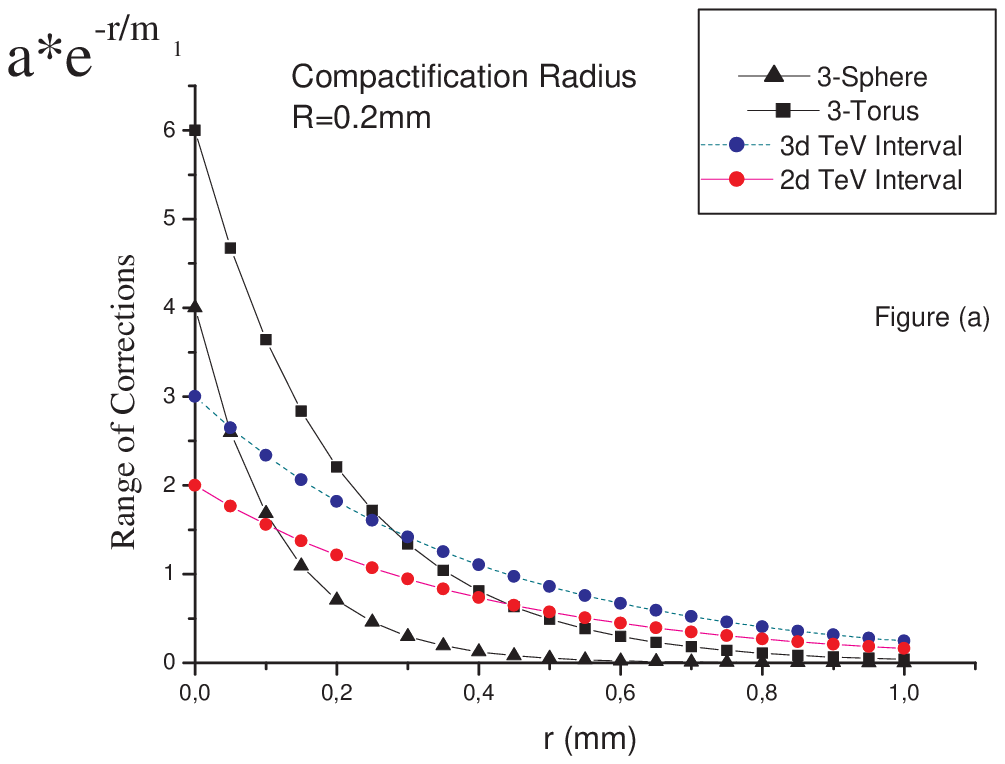}
\end{minipage}
\begin{minipage}{18pc}\hspace{2pc}%
\includegraphics[width=20pc]{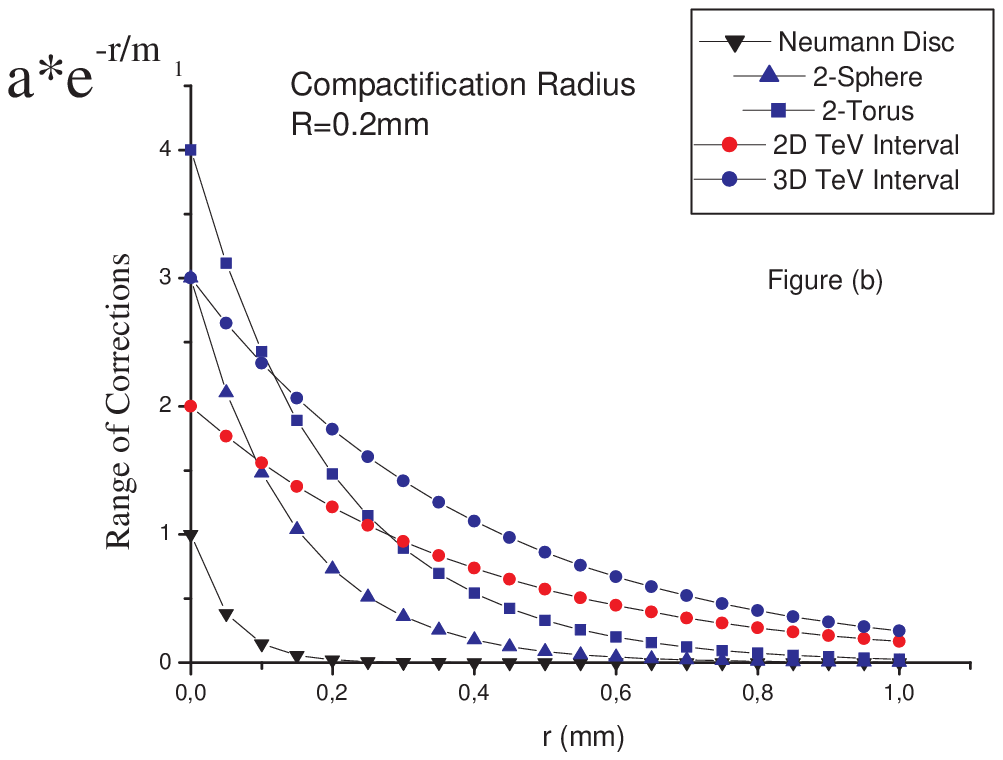}
\end{minipage}
\includegraphics[width=20pc]{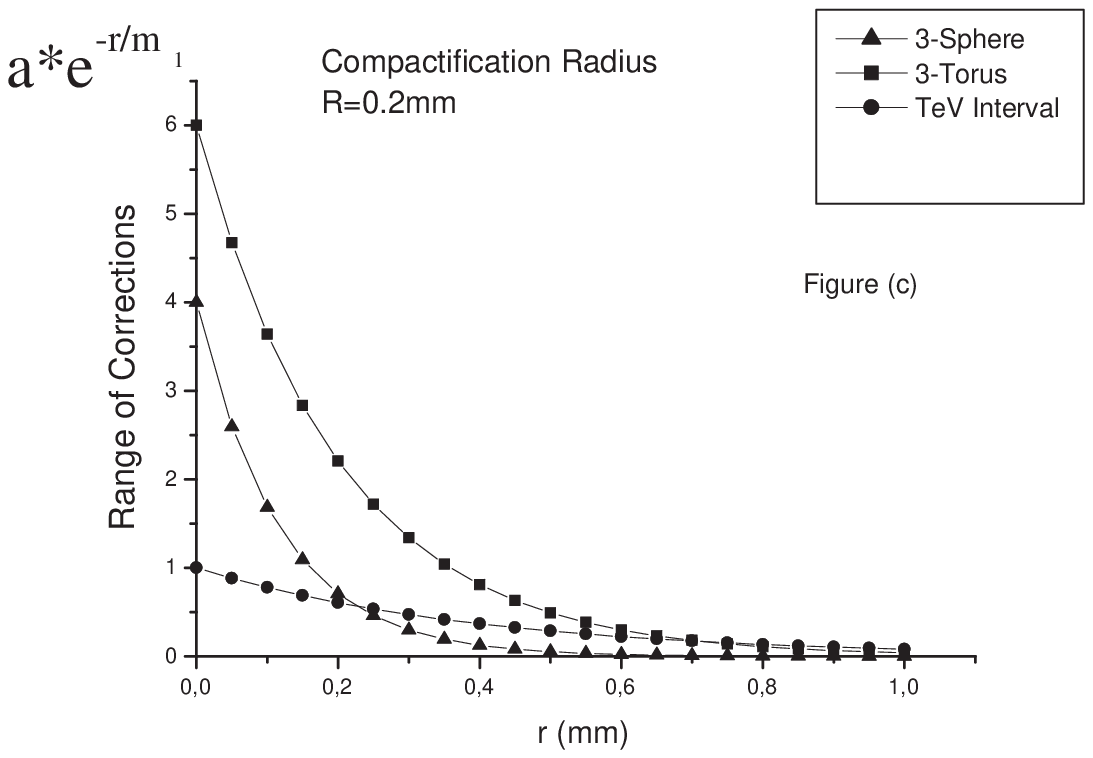}\hspace{0pc}%
\begin{minipage}[b]{18pc}\caption{Comparison of the 1d, 2d, 3d interval with Neumann boundaries,
 with 2-Torus, 3-Torus, 2-Sphere, 3-Sphere and the Neumann disk. The compactification radius is equal to $R = 0.2 {\,}mm$. The 2-dimensional spaces
 are shown in Figure (b) and the 3-dimensional spaces
 are shown in Figure (a). The shadowing effects are clearly shown in Figures (a), (b) and (c). In Figure (c) the shadowing effect occurs even for spaces with different spatial dimensions, namely two 3-dimensional with a 1-dimensional one.}
\label{neumtor}
\end{minipage}
\end{figure}
First of all, we compare the behavior of the gravitational potential along the physically-acceptable spaces we found
(the Neumann disk and the Neumann interval) to those
admitted by a 2-dimensional torus, $T^2$, and a 2-dimensional
sphere, $S^2$, as well as their three-dimensional counterparts, the
3-sphere and the 3-torus. Assuming that the torus radii are equal,
and keeping the lowest Kaluza-Klein mode (i.e., to leading-order correction
of the Newtonian potential), the gravitational potential corresponding
to the n-torus is given by \cite{Kehagias2001},
\begin{equation}\label{26} V_3 (r) = -\frac{G_N M}{r} \left ( 1 +
2n e^{- \frac{r}{R}} \right ) \: ,
\end{equation} while, as far as the n-sphere is concerned, the result is quite similar \cite{Kehagias2001},
\begin{equation}\label{26a} V_3 (r) = -\frac{G_N M}{r} \left ( 1 +
(n+1) e^{- \frac{\sqrt{n}r}{R}} \right ) \: . \end{equation}

\noindent In Figs. 1 and 2, we plot the Yukawa-type
correction-term, $a e^{- r/m_1 }$, versus $r$, corresponding to
the 1d, 2d, 3d interval with Neumann boundaries, as well as to the 2-Torus, 3-Torus, 2-Sphere, 3-Sphere and the Neumann
disk. In Fig. 1 the compactification radius is equal to $R = 0.2
{\,}mm$, while in Fig. 2 the compactification radius is equal to
$R = 0.05 \, mm$. These two values are quite near to the 2004 and
2006 experimental bounds, that is $0.2 \, mm$ and $0.056{\,}$mm,
respectively.

From Fig. 1 it becomes evident that, the corrections to the
gravitational potential due to the 2-Torus and the 2-Sphere are
larger, than those of the Neumann disk. This result also holds if
the compactification radius is smaller, for example $R = 0.05
{\,}mm$, see Fig. 2 (b).
\begin{figure}[ht!]
\begin{minipage}{18pc}
\includegraphics[width=20pc]{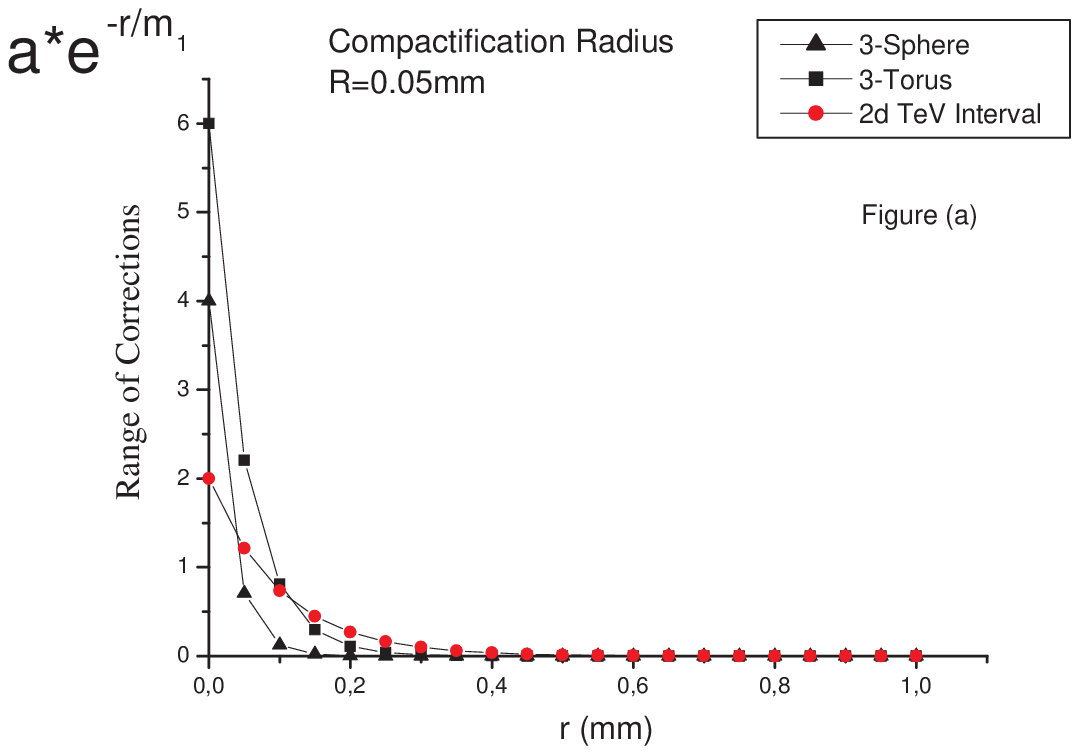}
\end{minipage}
\begin{minipage}{18pc}\hspace{2pc}%
\includegraphics[width=20pc]{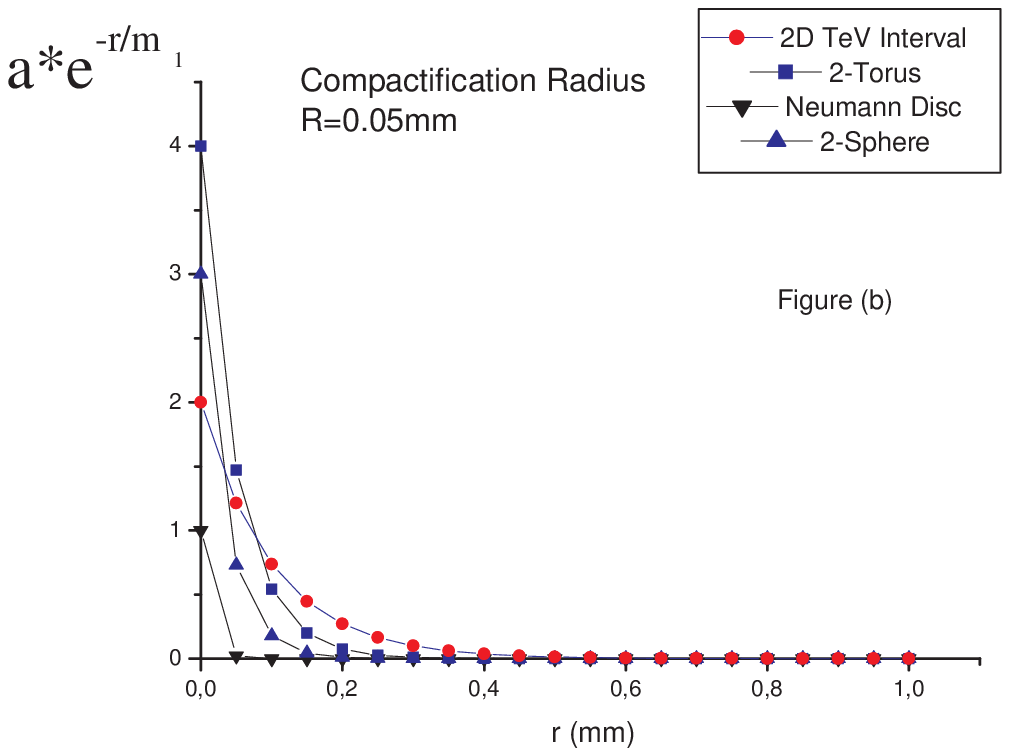}
\end{minipage}
\includegraphics[width=20pc]{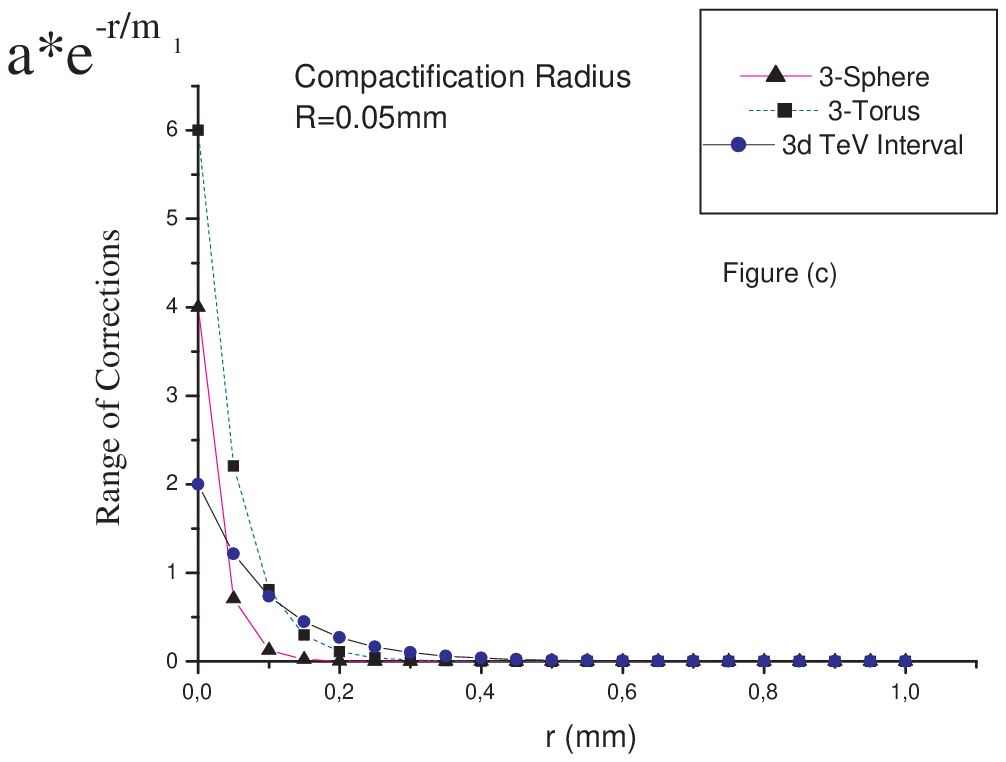}\hspace{0pc}%
\begin{minipage}[b]{18pc}\caption{Comparison of the 1d, 2d, 3d interval with Neumann boundaries,
 with 2-Torus, 3-Torus, 2-Sphere, 3-Sphere and the Neumann disk. The compactification radius is equal to $R = 0.05 {\,}mm$. The corrections are much smaller compared to the $R = 0.2 {\,}mm$ case. In Figs. (a) and (c) we compare the 3-sphere and the 3-torus with the two- and the three-dimensional interval, where the shadowing effect between spaces with different dimensions is clearly shown. In Fig. (b) we compare all the 2-dimensional spaces. We can see that the Neumann disk yields negligible corrections when compared to the other spaces.}
\label{neumtor}
\end{minipage}
\end{figure}
Observing Fig. 1 (a), which is the plot of the Yukawa corrections
corresponding to the 3-Sphere, 3-Torus, the 2d and 3d interval, we
can see that the N-N 2d and the corresponding 3d interval's
corrections can become smaller, equal (at some point) and larger
than the corrections corresponding to the torus- and the
sphere-case. This also holds for the 1d interval (see Fig. 1 (c)).
Such a behavior could arise problems, since we cannot figure out
the space that corresponds to the measured corrections, reflecting
an interesting phenomenon (often appearing in extra-dimensional
theories) known as shadowing \cite{Dienes2002,Oikonomou2011}.
Similar results are obtained when $R = 0.05 {\,}mm$, but the
corrections are quite smaller than those corresponding to the $R =
0.2 {\,}mm$ case. We must note that the plots appearing in Fig. 1
are for illustrative purposes only, since the value of the
compactification radius, namely $R = 0.2 {\,}mm$, is only
compatible to the 2004 experimental results, but excluded by the
2006 experiment. In addition the plots for the interval in Fig. 2
are also for illustrative purposes, since the maximum allowed
value for the radius corresponding to the interval space is
smaller than $R = 0.05 {\,}mm$ (see the list below and Table 1).

In view of Figs. 1 and 2, we need to stress that, if the
compactification radius is small, the corrections near the current experimental limit are
negligible. Nevertheless, we expect that, if the experiments
allow us to check Newton's law at smaller distances, the
corrections might be significant, as it is indicated also by \cite{Hoyle2004} and \cite{Hoyle2006}.

Before closing this section, we must address another important
issue that has to do with the experimental verification of the
spaces we described above. The experiments verify Newton's law
down to $0.056{\,}mm$, with $95\%$ confidence level
\cite{Hoyle2004,Hoyle2006}. In this case, the experimental
constraints are quite restrictive since \cite{Hoyle2006}:

\begin{itemize}

\item For one extra dimension, the radius $R$ must be smaller than
$0.044{\,}mm$, \linebreak that is $R\leq 44{\,}\mu m$

\item For two equal-size extra dimensions, the common radius $R$ must
be smaller than $0.0374{\,}mm$, that is $R\leq 37{\,}\mu m$

\item For $a\geq 1$, all radii of compactification with $m_1\geq 0.056{\,}mm$
are excluded.

\end{itemize}

We can estimate the size of each compact space we
studied, in order that the theoretical results to be
consistent with the experimental constraint $m_1\geq 0.056{\,}mm$.
In Table 1 we present the predictions for the radii of each
compact space.

\begin{table}
\begin{center}
\begin{tabular}{|c|c|}
  \hline
  \bf{Extra Dimensional Space} & \bf{Radius, R} \\
  \hline
  3-Torus $T^3$ & R=0.056 $mm$ \\
  \hline
  2-Torus $T^2$ & R=0.056 $mm$ \\
  \hline
  2-Sphere $S^2$ & R=0.079 $mm$ \\
  \hline
  3-Sphere $S^3$ & R=0.096 $mm$ \\
  \hline
  Neumann disk & R=0.2145 $mm$ \\
   \hline
  1-Dimensional Interval & R=0.028 $mm$ \\
   \hline
  2-Dimensional Interval & R=0.028 $mm$ \\
   \hline
  3-Dimensional Interval & R=0.028 $mm$ \\
  \hline
\end{tabular}
\end{center}\caption{The Radii of the compact spaces consistent with the experimental constraints on $m_1$}\label{table1}
\end{table}
The only results which are not consistent with the experimental
constraints are the 2-sphere and the Neumann disk. Therefore both
these extra spaces must be excluded.

\section{Conclusions}

In this article we studied possible modifications of Newton's law,
due to the existence of a compact extra-dimensional space, at the
boundary of which the extra-dimensional component of the
gravitational field obeys Dirichlet, Neumann or combination of
these boundary conditions. Usually, such a modification implements
Yukawa-type corrections, $\frac{a}{r} e^{- r/m_1}$, in addition to
the $\frac{1}{r}$ gravitational potential. We considered two types
of extra spaces, namely the disk of radius $R$ (two-dimensional)
and the compact interval, $I = [0, 2 \pi R]$ in one, two and
three dimensions. In the Dirichlet-disk case, the results
obtained are rather unexpected, i.e., the gravitational potential
in three dimensions takes the form of a Yukawa-type potential and
not of a Yukawa-type correction. Clearly, such a behavior is quite
far from physical reality, since the well-established Newton's
law, $\frac{1}{r}$, cannot be obtained. This result suggests that extra-dimensional models employing Dirichlet disks as extra spaces must be ruled out. On the contrary, an internal disk with
Neumann boundaries does predict Yukawa-type corrections to the
Newtonian potential, although, quantitatively, they are less
significant than those arising from other two-dimensional spaces (cf. Figs. 1 and 2).
However, this type of extra space must also be excluded, since the corresponding
corrections to Newton's law do not compromise the recent experimental
results (the same is also true for the 2-sphere).

There is another extra-dimensional spacetime which yields corrections to
the three-dimensional gravitational potential, in which the extra space corresponds to the
interval $I = [0, 2 \pi R]$. In connection to the
one-dimensional interval, we also studied the corresponding two- and three-dimensional intervals. We examined all the possible boundary
conditions that were used in the extra-dimensional
phenomenological models, using the interval as an extra space. We proved that the only acceptable combination (in
reference to a correct behavior of the Newton's law corrections)
is the interval with Neumann boundaries, i.e., our results rule out the
Dirichlet boundaries and combinations between Dirichlet and Neumann
boundaries.

Finally, let us note that the interval as an extra
dimensional space, is the simplest structure that an extra space
can have. Indeed, using the interval as an extra dimensional space,
we come across only to the problem of explaining how the extra
dimensions became so small and stabilized. When one considers for
example the torus, an additional problem must be dealt with, which is
the question why the extra space topology is non-trivial, but the
other visible three have trivial topology\footnote{Nevertheless,
we must note, that even the three visible dimensions of our world
may have non-trivial topology. This exotic scenario is based on
the assumption that the Universe topology can be different from
Euclidean beyond the present day particle horizon
\cite{DeOliveira1996}.}. This makes the interval quite
appealing as an extra space.

{\bf Acknowledgements:} Financial support by the Research
Committee of the Technological Education Institute of Serres,
under grant SAT/ME/260111-01/02, is gratefully acknowledged.

\end{document}